\documentclass[letterpaper,journal]{IEEEtran}
\usepackage[T1]{fontenc}
\usepackage{amsmath,amsfonts,amssymb,bm}
\usepackage{graphicx}
\usepackage{placeins}
\usepackage{cite}
\usepackage{url}
\usepackage{textcomp}
\usepackage[hidelinks]{hyperref}
\newtheorem{theorem}{Theorem}
\newtheorem{lemma}{Lemma}
\newtheorem{proposition}{Proposition}
\newtheorem{corollary}{Corollary}
\newtheorem{definition}{Definition}
\newcommand{\jj}{\mathrm{j}}
\newcommand{\Rea}{\operatorname{Re}}
\newcommand{\Ima}{\operatorname{Im}}
\newcommand{\diag}{\operatorname{diag}}
\newcommand{\adj}{\operatorname{adj}}
\newcommand{\C}{\mathbb C}
\newcommand{\V}{\mathcal V}
\newcommand{\E}{\mathcal E}
\newcommand{\I}{\mathcal I}

\begin{document}

\title{Feasibility of Pareto-Improving Damping Control of Forced Oscillations: The Active-Power Case}

\author{
	Haozong Wang,
	Yuru Wu, ~\IEEEmembership{Member,~IEEE,}
	Yan Wen, ~\IEEEmembership{Member,~IEEE,}
	Qian Liu, ~\IEEEmembership{Student Member,~IEEE,} \\
	Jiahui Yang, ~\IEEEmembership{Student Member,~IEEE,}
	Yayu Yang, ~\IEEEmembership{Member,~IEEE,} \\
	Muhammad Umar Afzaal, ~\IEEEmembership{Student Member,~IEEE,}
	and Yilu Liu, ~\IEEEmembership{Fellow,~IEEE}
	\thanks{Received xx, 2026; revised xx, 2026. \textit{(Corresponding author: Yuru Wu)}}
	\thanks{Haozong Wang, Yuru Wu, Yan Wen, Qian Liu, Jiahui Yang, Yayu Yang, Muhammad Umar Afzaal, and Yilu Liu are with with the Department of Electrical Engineering and Computer Science, University of Tennessee, Knoxville, TN 37996, USA. Yuru Wu and Yilu Liu are also with the Oak Ridge National Laboratory, Oak Ridge, TN 37831, USA. (e-mail: hwang124@utk.edu, ywu70@vols.utk.edu, ywen7@utk.edu, qliu30@vols.utk.edu, jyang75@vols.utk.edu, yyang117@utk.edu, mafzaal2@utk.edu, liu@utk.edu)}}

\markboth{Journal of \LaTeX\ Class Files,~Vol.~14, No.~8, August~2021}%
{Shell \MakeLowercase{\textit{et al.}}: A Sample Article Using IEEEtran.cls for IEEE Journals}

\IEEEpubid{0000--0000/00\$00.00~\copyright~2021 IEEE}

\maketitle

\begin{abstract}
Damping control of a forced oscillation may reduce amplitudes at certain locations while increasing them elsewhere. This paper introduces Pareto improvement as a criterion for assessing such control: at least one nodal oscillation amplitude decreases, while none increases across all other nodes. The analysis considers active-power forcing sources and active-power control injections in connected swing networks with positive inertia and damping. Necessary conditions for Pareto improvement feasibility are derived from the normalized nodal responses. It is proven that for every network with a single oscillation source and a single actuator, at least one positive forcing frequency is shown to preclude Pareto improvement, regardless of the amplitude and phase of the control input. A three-node counterexample shows that this limitation does not persist with multiple actuators. PSS/E simulations of a 6-bus benchmark system and a modified IEEE 39-bus system validated the analysis. These findings suggest the value of assessing control effects across the network and further investigating multi-actuator damping control.
\end{abstract}

\section{Introduction}\label{sec:intro}
\IEEEPARstart{O}{scillations} are a persistent concern for the secure operation of interconnected power systems. Large or sustained oscillations can impose repeated mechanical stress on equipment, degrade power quality, trigger unintended protection actions, and restrict power transfers~\cite{nerc2017,zhu2021}. Their effects can extend far beyond the equipment or area where they originate. Field events documented by the North American Electric Reliability Corporation (NERC) demonstrate that a localized disturbance can produce substantial power swings across an interconnection~\cite{nerc2017}. Understanding the mechanisms that sustain these oscillations and establishing effective control measures are therefore important for system operation.

Oscillations can be categorized into natural and forced types based on their distinct mechanisms~\cite{pes2023,ghorbaniparvar2017}. In small-signal analysis, natural oscillations reflect the system's intrinsic modes: their frequencies and damping ratios are determined by the eigenvalues of the linearized dynamics. Forced oscillations, in contrast, are sustained by a persistent periodic disturbance and can occur in an otherwise well-damped system. The disturbance determines the forcing frequency, while the network shapes the response magnitude and spatial distribution. Studies have shown that resonant amplification can occur when the forcing frequency is close to a natural modal frequency~\cite{sarmadi2015,ye2017}. Because the driving input persists, locating and disconnecting the source is widely regarded as the most effective way to eliminate a forced oscillation~\cite{chevalier2019}. NERC guidance similarly prioritizes identifying and disabling the forcing input, either by disconnecting the responsible equipment or by correcting its operation~\cite{nerc2021}.

The need to eliminate forcing sources has motivated extensive research on forced oscillation source localization. Accurate localization requires separating the source from other devices that respond strongly to the same disturbance. Measurement-based methods provide the necessary information about the oscillatory components: Follum \emph{et al.}~\cite{follum2017} jointly estimated electromechanical modes and forced oscillations from measured signals. For source location, Chen \emph{et al.}~\cite{chen2013} related energy dissipation to generator damping and proposed an energy-based method that traces the source through the network. Building on this work, Maslennikov \emph{et al.}~\cite{maslennikov2017} developed the dissipating energy flow method with signal processing tailored to practical phasor measurement unit (PMU) data, enabling automated analysis of measured oscillation events. Yu \emph{et al.}~\cite{yu2024} estimated energy flows over virtual connections between monitored buses and used the resulting dissipating potentials to locate forcing sources with partial synchrophasor coverage, without requiring network topology information. Chevalier \emph{et al.}~\cite{chevalier2019} used Bayesian inference to account for uncertainty in generator parameters, and Delabays \emph{et al.}~\cite{delabays2023} developed a maximum-likelihood approach that infers forcing location and frequency together with dynamical model information. These methods connect measurements of a distributed response to corrective action at its origin.

\IEEEpubidadjcol

Despite the emphasis on source removal, a substantial body of work also investigates forced oscillation damping control. Such control schemes are expected to reduce the disturbance impact while the source is being located or before it can be disconnected~\cite{trudnowski2020,zhu2021}. Trudnowski and Guttromson~\cite{trudnowski2020} proposed tuned feedback that uses controllable devices to generate a compensating oscillation. Surinkaew \emph{et al.}~\cite{surinkaew2021} designed a power oscillation damper to address both inter-area and forced oscillations. Zhu \emph{et al.}~\cite{zhu2021} combined source location with active-power modulation of battery energy storage systems. Wilches-Bernal \emph{et al.}~\cite{wilches2020} examined forced responses under a wide-area damping controller acting through the Pacific DC Intertie. Zhao \emph{et al.}~\cite{zhao2021} used wind farms under grid-following and grid-forming control to isolate and suppress oscillation propagation. Together, these studies demonstrate several ways to attenuate forced responses through controlled power injections.

These control schemes motivate a more fundamental examination of when forced oscillation damping control can achieve the intended response reduction. Two considerations are particularly relevant. First, growing use of power electronic devices introduces additional control dynamics and extends the frequency ranges relevant to forced oscillation analysis~\cite{hatziargyriou2021}. Forced excitation also takes new forms: Chen \emph{et al.}~\cite{chen2025} showed that interharmonics produced by maximum power point tracking in a grid-following inverter can drive oscillations in grid-forming inverters. The diversity of excitation mechanisms and time scales calls for re-examining the assumptions and bandwidths underlying existing damping control strategies. Second, a forced oscillation is maintained by continuing energy injection from its source, with a frequency that need not coincide with a natural mode. System modes influence its amplification, but the response also depends on where and how the disturbance enters the network~\cite{sarmadi2015,ye2017}. The control input must counter the response sustained by this ongoing input. In a linear model, its contribution combines with the source-driven response at each node, making attenuation depend on their relative amplitudes and phases. Consequently, improving modal damping alone does not guarantee that forced response will decrease. This difference between accelerating the decay of a natural response and opposing a continuously driven response motivates studying feasibility before choosing a control law or tuning its parameters.

In this paper, we introduce \emph{Pareto improvement} as a criterion for evaluating forced oscillation damping strategies. A strategy is Pareto-improving if, compared to the baseline response, it reduces at least one nodal oscillation amplitude without increasing any of the others. We study this criterion using a small-signal model based on the swing equations~\cite{bergen1981}, with both the disturbance and the control inputs represented as active-power injections. The analysis compares single-actuator control with multi-actuator control using coordinated inputs. PSS/E simulations examine the results using a 6-bus benchmark system and a modified 39-bus system with synchronous generators (SGs) and virtual synchronous machines (VSMs). The benchmark system also demonstrates amplitude reductions under multi-actuator control. The main contributions are as follows.
\begin{enumerate}
\item Necessary conditions for Pareto improvement are derived at a prescribed frequency. After accounting for zero baseline responses, the normalized control responses at affected dynamic nodes must lie in a common open half-plane.
\item A universal single-actuator limitation is proved: every fixed network satisfying the assumptions and every distinct source--actuator pair have at least one positive frequency at which Pareto improvement is infeasible, even with unrestricted control amplitude and phase.
\item Multi-actuator control can overcome the single-actuator limitation. A three-node counterexample shows that two coordinated injections can cancel one node's response at every positive frequency while leaving the other two responses unchanged.
\end{enumerate}

The remainder of this paper is organized as follows. Section~\ref{sec:model} introduces the model and Pareto improvement criterion. Section~\ref{sec:geometry} derives necessary conditions for Pareto improvement, and Section~\ref{sec:limit} proves the single-actuator limitation. Section~\ref{sec:multi} presents the multi-actuator counterexample, and Section~\ref{sec:studies} reports the simulation study. Section~\ref{sec:discussion} addresses practical implications, limitations, and future work. Section~\ref{sec:conclusion} concludes the paper. 

\section{Model and Pareto Improvement Criterion}\label{sec:model}

\subsection{Grid Model}

We adopt Kron reduction \cite{dorfler2013} of power grid in this work, which retains nodes with dynamic equations and eliminates nodes described only by algebraic equations. Let $\mathcal G=(\V,\E)$ denote the reduced network with $n\geq2$ dynamic nodes, each following the swing-equation dynamics. The edges represent equivalent electrical couplings and need not correspond to individual physical lines.

Let $\delta_i$ be the voltage angle, and let $P_{e,i}(\delta)$ denote the active power transferred from dynamic node $i$ to the network. In the classical lossless network representation~\cite{dorfler2013}, with voltage magnitudes $E_i>0$ and coupling reactances $x_{ik}=x_{ki}>0$, this power is given by:
\begin{equation}\label{eq:electrical_power}
 P_{e,i}(\delta)=\sum_{k:\{i,k\}\in\E}\frac{E_iE_k}{x_{ik}}
 \sin(\delta_i-\delta_k)
\end{equation}

The active-power balance at each dynamic node is described by the swing equation~\cite{bergen1981}:
\begin{equation}\label{eq:nodal_balance}
 m_i\ddot\delta_i+d_i\dot\delta_i
 =P_i^0+\Delta P_i(t)-P_{e,i}(\delta)
\end{equation}

where $m_i>0$ and $d_i>0$ are the inertia and damping coefficients of the represented device, $P_i^0$ is its steady mechanical input and $\Delta P_i$ is an additional active power injection.

At a synchronous operating point $\delta^0$, the angles are constant and $P_i^0=P_{e,i}(\delta^0)$. Define the angle deviation $\theta_i=\delta_i-\delta_i^0$. Then equation \eqref{eq:nodal_balance} could be re-written as:
\begin{equation}\label{eq:deviation_balance}
 m_i\ddot\theta_i+d_i\dot\theta_i
 =\Delta P_i(t)-\bigl[P_{e,i}(\delta^0+\theta)-P_{e,i}(\delta^0)\bigr]
\end{equation}

The first-order change in the electrical power follows by expanding each sine term in~\eqref{eq:electrical_power}:
\begin{equation}\label{eq:power_linearization}
 \begin{aligned}
 P_{e,i}(\delta^0+\theta)-P_{e,i}(\delta^0)
 &=\sum_{k:\{i,k\}\in\E}w_{ik}(\theta_i-\theta_k)\\
 &\quad+O(\|\theta\|^2)
 \end{aligned}
\end{equation}

where $w_{ik}$ is the derivative of the power-angle relation for edge $\{i,k\}$ at the operating point:
\begin{equation}\label{eq:synchronizing_weight}
 w_{ik}=w_{ki}=\frac{E_iE_k}{x_{ik}}\cos(\delta_i^0-\delta_k^0)
\end{equation}

In the stable operating region, the angle difference has the boundary $|\delta_i^0-\delta_k^0|<\pi/2$, so~\eqref{eq:synchronizing_weight} gives $w_{ik}>0$. Set $w_{ik}=0$ when the edge is absent. Define the weighted Laplacian by
\begin{equation}\label{eq:L}
 L_{ik}=-w_{ik}\ (i\neq k),\qquad L_{ii}=\sum_{k\neq i}w_{ik}
\end{equation}

Neglecting the higher-order terms, then we can obtain the linearized form of equation \eqref{eq:power_linearization}:
\begin{equation}
	(L\theta)_i=\sum_{k\neq i}w_{ik}(\theta_i-\theta_k)
\end{equation}

Expressing inertia and damping in matrix form yields:
\begin{equation}\label{eq:MD}
 M=\diag(m_i)_{i=1}^n\succ0,\qquad D=\diag(d_i)_{i=1}^n\succ0
\end{equation}

Considering a disturbance $q(t)$ at dynamic node $s$ and a control injection $u(t)$ from an actuator at a distinct dynamic node $c$, the model becomes
\begin{equation}\label{eq:swing}
 M\ddot\theta+D\dot\theta+L\theta=e_s q(t)+e_c u(t),\quad s\neq c
\end{equation}
where $e_i$ is the $i$th coordinate vector. Note that equation~\eqref{eq:swing} is the same form as the model in~\cite{poolla2017}, whose nonlinear swing-network counterpart is described in~\cite{dorfler2013}.

Equation~\eqref{eq:swing} describes the dynamic nodes. To assess all nodes, we also recover the angles of the eliminated nodes. Let $\mathcal A$ denote the eliminated algebraic nodes, representing network buses removed by Kron reduction. Partition the full network Laplacian $\mathcal L$ by dynamic nodes $\V$ and eliminated nodes $\mathcal A$. With active power injections at the eliminated nodes fixed at their steady-state values, the angle deviations satisfy
\begin{equation}\label{eq:full_network_balance}
 \begin{aligned}
 M\ddot\theta+D\dot\theta+\mathcal L_{\V\V}\theta
       +\mathcal L_{\V\mathcal A}\theta_{\mathcal A}
       &=e_s q+e_c u\\
 \mathcal L_{\mathcal A\V}\theta
       +\mathcal L_{\mathcal A\mathcal A}\theta_{\mathcal A}&=0
 \end{aligned}
\end{equation}

Eliminating the second equation gives
\begin{equation}\label{eq:network_reduction}
 \begin{aligned}
 \theta_{\mathcal A}&=T\theta,\qquad
 T=-\mathcal L_{\mathcal A\mathcal A}^{-1}\mathcal L_{\mathcal A\V},\\
 L&=\mathcal L_{\V\V}
     -\mathcal L_{\V\mathcal A}\mathcal L_{\mathcal A\mathcal A}^{-1}
        \mathcal L_{\mathcal A\V}
 \end{aligned}
\end{equation}

For a connected network with positive edge weights, the grounded block $\mathcal L_{\mathcal A\mathcal A}$ is nonsingular, and its Schur complement $L$ is again a connected weighted Laplacian~\cite{dorfler2013}. The matrix $T$ recovers the angles of the eliminated nodes from those of the dynamic nodes.

\subsection{Frequency Response and Network Properties}

At a prescribed angular frequency $\omega>0$, write
\begin{equation}\label{eq:phasors}
 q(t)=\Rea\{\widehat q e^{\jj\omega t}\},\qquad
 u(t)=\Rea\{Ue^{\jj\omega t}\},\qquad \widehat q\neq0
\end{equation}

Here $\jj^2=-1$ and $U\in\C$. Let \(p\in\mathbb C\) denote the Laplace variable, define
\begin{equation}\label{eq:QH}
 Q(p)=Mp^2+Dp+L,\qquad H(p)=Q(p)^{-1}
\end{equation}
Let $h_{ik}(p)$ denote the entry in row $i$ and column $k$ of $H(p)$.

For a set $\mathcal J$ of dynamic nodes, $Q_{\mathcal J\mathcal J}$ denotes the principal submatrix indexed by $\mathcal J$. Superscript $*$ denotes conjugate transpose.

The following lemma establishes the existence of the frequency response and the matrix properties used in the later analysis.

\begin{lemma}\label{lem:invertible}
For every $\omega>0$, $Q(\jj\omega)$ and every nonempty principal submatrix $Q_{\mathcal J\mathcal J}(\jj\omega)$ are nonsingular. Moreover,
\begin{equation}\label{eq:imag}
 \frac{H(\jj\omega)-H(\jj\omega)^*}{2\jj}
 =-\omega H(\jj\omega)^*D H(\jj\omega)\prec0
\end{equation}

In particular, $h_{ii}(\jj\omega)\neq0$ for every $i$, and $h_{ik}=h_{ki}$.
\end{lemma}

\begin{IEEEproof}
If $Q_{\mathcal J\mathcal J}(\jj\omega)x=0$, then
\begin{equation}
 0=\Ima\{x^*Q_{\mathcal J\mathcal J}(\jj\omega)x\}=\omega x^*D_{\mathcal J\mathcal J}x
\end{equation}

Since $D_{\mathcal J\mathcal J}\succ0$, this implies $x=0$. The argument also applies to $\mathcal J=\V$. Further, $Q-Q^*=2\jj\omega D$, and the following relation could be derived:
\begin{equation}
 H-H^*=H^*(Q^*-Q)H=-2\jj\omega H^*DH
\end{equation}

As $H$ is invertible and $D\succ0$, division by $2\jj$ gives the negative definite matrix in~\eqref{eq:imag}. Its $i$th diagonal entry is $\Ima h_{ii}<0$, establishing the nonzero diagonal property. Finally, $Q^{\mathsf T}=Q$ implies $H^{\mathsf T}=H$.
\end{IEEEproof}

Let $Y\in\C^n$ collect the complex angle amplitudes at the dynamic nodes. The magnitude and phase of $Y_i$ give those of the oscillation at dynamic node $i$. Substituting $\theta(t)=\Rea\{Y e^{\jj\omega t}\}$ into~\eqref{eq:swing} gives
\begin{equation}\label{eq:harmonic_balance}
 Q(\jj\omega)Y=e_s\widehat q+e_cU
\end{equation}

At each $\omega>0$, Lemma~\ref{lem:invertible} guarantees that $Q(\jj\omega)$ is nonsingular. Hence, for a specified source phasor $\widehat q$ and control phasor $U$, \eqref{eq:harmonic_balance} uniquely determines the response vector $Y$. Define
\begin{equation}\label{eq:ab}
 a=H(\jj\omega)e_s\widehat q,\qquad b=H(\jj\omega)e_c
\end{equation}

Here $a$ is the baseline angle response to the source alone. Each $b_i$ is the complex gain from active power injected at dynamic node $c$ to the angle at dynamic node $i$. The control contributes $bU$, so linearity gives the total response
\begin{equation}\label{eq:Y}
 Y=a+bU
\end{equation}

\subsection{Pareto Improvement in Nodal Amplitudes}
Let $\mathcal N=\V\cup\mathcal A$ include all dynamic and eliminated nodes. Listing dynamic nodes first gives the angle-output vector $y_{\rm out}=C\theta$. Its complex oscillation amplitudes satisfy
\begin{equation}\label{eq:full_output_response}
 \begin{aligned}
 C&=\begin{bmatrix}I\\T\end{bmatrix},\qquad
 \widetilde a=Ca,\quad \widetilde b=Cb,\\
 \widetilde Y&=CY=\widetilde a+\widetilde bU
 \end{aligned}
\end{equation}

The identity block includes every dynamic-node response, and $T$ recovers the responses at the eliminated nodes.

\begin{definition}[Pareto improvement]\label{def:improve}
On the node set $\mathcal N$, an injection $U$ produces a \emph{Pareto improvement} at $\omega$ if
\begin{equation}\label{eq:target}
 \begin{aligned}
 |\widetilde Y_i|&\leq|\widetilde a_i|\quad\text{for all }i\in\mathcal N,\\
 |\widetilde Y_k|&<|\widetilde a_k|\quad\text{for some }k\in\mathcal N
 \end{aligned}
\end{equation}
\end{definition}

This definition is motivated by the use of selected measurements in some practical damping control schemes. Many damping control schemes use local or remote feedback signals to reduce the oscillations observed at those locations~\cite{wilches2020,zhu2021,trudnowski2020}. For natural oscillations without a continuing excitation, this scheme works fine because the decrease in the observed oscillation implies energy dissipation. For forced oscillations, however, the source remains active. A smaller measured amplitude may reflect local cancellation between the source and control responses, while neglects the possibility that amplitudes may increase at other nodes. We therefore want to assess the benefit of control across the full network, including nodes not used for feedback. Pareto improvement formalizes this requirement: control must not make the oscillation worse at any other node.

\section{Necessary Conditions for Pareto Improvement at a Prescribed Frequency}\label{sec:geometry}

This section gives necessary conditions for Pareto improvement. First, consider two zero-response cases at dynamic nodes. If $a_i=0$ but $b_i\neq0$, any nonzero input creates an oscillation at dynamic node $i$. Preserving its zero baseline requires $U=0$, ruling out Pareto improvement. If $b_i=0$, control leaves the response unchanged, so the amplitude constraint is automatically satisfied. After excluding the first case, we focus on dynamic nodes with $b_i\neq0$, whose baseline responses are nonzero. Denote this set by $\mathcal I$.

\begin{equation}\label{eq:I}
 \I=\{i\in\V:b_i\neq0\}
\end{equation}

This set is nonempty because $b_c=h_{cc}\neq0$. If no dynamic node has a zero-baseline obstruction, define the nonzero ratios
\begin{equation}\label{eq:r}
 r_i=b_i/a_i,\qquad i\in\I
\end{equation}

\begin{theorem}[Necessary conditions for Pareto improvement]\label{thm:geometry}
Pareto improvement on $\mathcal N$ at a prescribed $\omega>0$ requires both of the following conditions:
\begin{enumerate}
\item There is no dynamic node with $a_i=0$ and $b_i\neq0$.

\item There exists $U_{\mathrm{dir}}\in\C$ such that
\begin{equation}\label{eq:direction}
 \Rea(r_iU_{\mathrm{dir}})<0\qquad\text{for every }i\in\I
\end{equation}
\end{enumerate}

Geometrically, the second condition is also equivalent to the fact that, in the complex plane, all $r_i$, $i\in\I$, lie in a common open half-plane whose boundary passes through the origin.
\end{theorem}

\begin{IEEEproof}
The first necessary condition was proved in the first paragraph of this section. Next we give the proof of the second necessary condition. Obviously, a Pareto improvement requires $U\neq0$. It also requires
\begin{equation}\label{eq:condition0}
|a_i+b_iU|\leq|a_i|
\end{equation}

Divide \eqref{eq:condition0} by $|a_i|$ and expand:
\begin{equation}\label{eq:quadratic}
 |1+r_iU|^2-1=2\Rea(r_iU)+|r_iU|^2\leq0
\end{equation}

Since $r_i\neq0$ and $U\neq0$, the quadratic term is strictly positive, so there should be $\Rea(r_iU)<0$ for every $i\in\I$. Taking $U_{\mathrm{dir}}=U$ proves the second necessary condition.
\end{IEEEproof}

\section{An Unavoidable Limit of Single Actuation}\label{sec:limit}

In this section, we prove that every single actuator located at a different dynamic node from the forced oscillation source encounters at least one frequency at which Pareto improvement is impossible. 

\begin{samepage}
\begin{theorem}[Universal single-actuator limitation]\label{thm:main}
Fix any grid satisfying~\eqref{eq:L}--\eqref{eq:swing}, and any distinct dynamic nodes $s,c$ for the source and actuator. There exists a finite $\omega_*>0$ such that
\begin{equation}\label{eq:onlyzero}
 |a_i+b_iU|\leq|a_i|\ \text{for all }i\in\V
 \quad\Longrightarrow\quad U=0
\end{equation}

Consequently, Pareto improvement on $\mathcal N$ is infeasible.
\end{theorem}
\end{samepage}

\subsection{Polynomial Facts Used in the Proof}
Let $N_{ik}(p)$ be the cofactor of the $(i, k)$ entry of $Q(p)$:
\begin{equation}\label{eq:cofactors}
 N_{ik}(p)=(-1)^{i+k}\det Q_{\V\setminus\{k\},\,\V\setminus\{i\}}(p)
\end{equation}

Here $\det$ denotes the determinant, and the submatrix is obtained by deleting row $k$ and column $i$ from $Q(p)$. Since the entries of $Q(p)$ are polynomials in $p$, each $N_{ik}(p)$ is a scalar polynomial.

Since $H(p)=Q(p)^{-1}$, the cofactor formula gives its $(i,k)$ entry as
\begin{equation}\label{eq:transfer_cofactor_relation}
 h_{ik}(p)=\frac{N_{ik}(p)}{\det Q(p)}
\end{equation}

Let $\ell\geq1$ be the number of edges on a shortest path from $s$ to $c$.
\begin{lemma}[Cofactor structure]\label{lem:cofactors}
The following properties hold:
\begin{enumerate}
\item $N_{ss}$ and $N_{cc}$ have degree $2(n-1)$, positive leading coefficients, and all roots in the open left half-plane.
\item $N_{sc}$ has degree $2(n-\ell-1)$ and a positive leading coefficient.
\item A real $\tau>0$ exists such that $N_{ik}(0)=\tau$ for every $i,k$.
\end{enumerate}
\end{lemma}

Appendix~\ref{app:cofactors} provides the proof.

\begin{lemma}[Continuous polynomial phase]\label{lem:phase}
Let $\psi$ be a real polynomial of degree $d_\psi$ with $\psi(0)>0$ and no imaginary-axis roots. Let $z_+(\psi)$ count its open-right-half-plane roots with multiplicity. The continuous phase $\phi_{\psi}$ of $\psi(\jj\omega)$, fixed by $\phi_{\psi}(0)=0$, satisfies
\begin{equation}\label{eq:phaselemma}
 \lim_{\omega\to\infty}\phi_{\psi}(\omega)
 =\left(d_\psi-2z_+(\psi)\right)\frac{\pi}{2}
\end{equation}
\end{lemma}

Appendix~\ref{app:phase} proves the formula. Note that the phase $\phi_{\psi}(\omega)$ is continuously unwrapped from its initial value at zero.

\subsection{Proof of Theorem~\ref{thm:main}}
\begin{IEEEproof}
There are two exhaustive cases.

\emph{Case 1.} Suppose $h_{sc}(\jj\omega_*)=0$ for some $\omega_*>0$. Thus
\begin{equation}
 a_c=h_{cs}(\jj\omega_*)\widehat q=0,\qquad
 b_c=h_{cc}(\jj\omega_*)\neq0
\end{equation}

This contradicts the first necessary condition in Lemma~\ref{lem:invertible}, proving~\eqref{eq:onlyzero}.

\emph{Case 2.} For any $\omega > 0$, $h_{sc}(j\omega) \neq 0$.
Define
\begin{equation}\label{eq:R}
 R(p)=\frac{h_{ss}(p)h_{cc}(p)}{h_{sc}(p)^2}
     =\frac{N_{ss}(p)N_{cc}(p)}{N_{sc}(p)^2}.
\end{equation}
All the matrix entries appearing here are nonzero on the positive imaginary axis. At dynamic nodes $s$ and $c$,
\begin{equation}\label{eq:rsource}
 r_s=\frac{h_{sc}}{h_{ss}\widehat q},\qquad
 r_c=\frac{h_{cc}}{h_{sc}\widehat q},\qquad
 \frac{r_c}{r_s}=R(\jj\omega)
\end{equation}

Consider when $R(\jj\omega)$ is a negative real number, $r_s$ and $r_c$ differ in phase by $180^\circ$. They cannot lie in a common open half-plane, so the necessary condition 2 in Lemma~\ref{lem:invertible} fails.

We now show that such a value must occur. In Case 2, $h_{sc}(\jj\omega)\neq0$ for every $\omega>0$. Since $Q(\jj\omega)$ is nonsingular, the relation $N_{sc}(\jj\omega)=h_{sc}(\jj\omega)\det Q(\jj\omega)$ gives $N_{sc}(\jj\omega)\neq0$. Thus $N_{sc}$ has no zeros on the positive imaginary axis. Its real coefficients exclude zeros on the negative imaginary axis as well because such a zero would have a conjugate on the positive imaginary axis. Finally, $N_{sc}(0)=\tau>0$ excludes a zero at the origin.

Let $z_+=z_+(N_{sc})$ count the roots of $N_{sc}$ in the open right half-plane, including multiplicity. The phase of $R$ is the sum of the phases of its two numerator polynomials minus twice the phase of its denominator polynomial. Accordingly, define
\begin{equation}\label{eq:Phi}
	\Phi(\omega)=\phi_{N_{ss}}(\omega)+\phi_{N_{cc}}(\omega)
	-2\phi_{N_{sc}}(\omega)
\end{equation}

All three polynomials are nonzero on the imaginary axis, so their phases can be followed continuously. At $\omega=0$, each polynomial equals $\tau>0$. Their phases therefore start at zero, giving $\Phi(0)=0$. 

By Lemma~\ref{lem:cofactors}, $N_{ss}$ and $N_{cc}$ each have degree $2(n-1)$ and all their roots in the open left half-plane, whereas $N_{sc}$ has degree $2(n-\ell-1)$. Let $z_+\geq0$ denote the number of roots of $N_{sc}$ in the open right half-plane, counted with multiplicity. Applying Lemma~\ref{lem:phase} to these three polynomials gives
\begin{align}\label{eq:totalphase}
	\lim_{\omega\to\infty}\Phi(\omega)
	&=2(n-1)\pi-2(n-\ell-1-z_+)\pi\notag\\
	&=2\pi(\ell+z_+)\geq2\pi
\end{align}

Thus the phase starts at zero and eventually exceeds $\pi$. Choose a finite $\Omega>0$ for which $\Phi(\Omega)>\pi$. Continuity then guarantees a frequency $\omega_\in(0,\Omega)$ at which $\Phi(\omega_*)=\pi$. At this frequency, $R(\jj\omega_*)$ has phase $\pi$ and a finite, nonzero magnitude, so it is real and negative.
Therefore, no nonzero $U$ can satisfy \eqref{eq:direction} when $r_s$ and $r_c$ have opposite phases. This proves~\eqref{eq:onlyzero}.
\end{IEEEproof}

\begin{corollary}[A conditional frequency count]\label{cor:count}
If $h_{sc}(\jj\omega)\neq0$ for every $\omega>0$, at least $\ell+z_+$ distinct positive frequencies are infeasible for Pareto improvement.
\end{corollary}

\begin{IEEEproof}
This follows by applying the above proof at the phases $(2k+1)\pi$, $k=0,\ldots,\ell+z_+-1$.
\end{IEEEproof}

Note that the limitation in Theorem~\ref{thm:main} is independent of the control law, since it holds for every complex amplitude $U$ of the sinusoidal control injection.

However, the infeasibility over the entire frequency band does not imply that a specific local frequency point or sub-band is infeasible. Appendix~\ref{app:full_node_control} gives a feasibility criterion and constructs a Pareto-improving injection when it holds.

\section{A Multi-Actuator Counterexample}\label{sec:multi}
A three-node counterexample shows that multi-actuator control with coordinated inputs can overcome the limitation in Theorem~\ref{thm:main}.

\begin{figure}[t]
\centering
\includegraphics{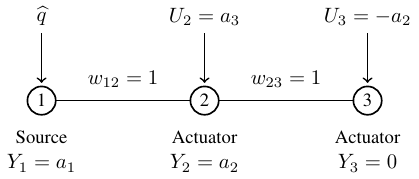}
\caption{Three-node counterexample for multi-actuator control.}
\label{fig:path}
\end{figure}

Consider Fig.~\ref{fig:path}, with unit inertia, damping, and edge weights. The three retained dynamic nodes form the full network, so $\mathcal N=\V=\{1,2,3\}$ and $\mathcal A$ is empty. The oscillation source is at dynamic node 1, and control inputs are available at dynamic nodes 2 and 3. The network matrix is
\begin{equation}\label{eq:pathQ}
 Q(p)=\begin{bmatrix}
 p^2+p+1&-1&0\\
 -1&p^2+p+2&-1\\
 0&-1&p^2+p+1
 \end{bmatrix}
\end{equation}
For a prescribed $\omega>0$, let $a=H(\jj\omega)e_1\widehat q=[a_1, a_2, a_3]^{\mathsf T}$, with $\widehat q\neq0$.

\begin{proposition}\label{prop:counter}
For~\eqref{eq:pathQ}, the injections
\begin{equation}\label{eq:pathcontrols}
 U_2=a_3,\qquad U_3=-a_2
\end{equation}
achieve
\begin{equation}\label{eq:pathY}
 Y_1=a_1,\qquad Y_2=a_2,\qquad Y_3=0
\end{equation}
at every positive frequency. Since $a_3\neq0$, these responses produce a Pareto improvement at every such frequency.
\end{proposition}

\begin{IEEEproof}
The third row of $Q(\jj\omega)a=e_1\widehat q$ gives
\begin{equation}\label{eq:pathrow}
 a_2=(1-\omega^2+\jj\omega)a_3
\end{equation}
Substituting $U_2=a_3$ and $U_3=-a_2$ into the controlled response yields
\begin{align}
 Y
 &=a+H(\jj\omega)(e_2a_3-e_3a_2)\notag\\
 &=a+a_3H(\jj\omega)
 \begin{bmatrix}
 0\\1\\-(1-\omega^2+\jj\omega)
 \end{bmatrix}\notag\\
 &=a-a_3H(\jj\omega)Q(\jj\omega)e_3\notag\\
 &=a-a_3e_3=[a_1,a_2,0]^{\mathsf T}
\end{align}
The third equality uses the third column of~\eqref{eq:pathQ}, and the last equality follows from $H(\jj\omega)Q(\jj\omega)=I$.

The $(1,3)$ cofactor of $Q(p)$ is 1, so
\begin{equation}\label{eq:a3}
 a_3=\frac{\widehat q}{\det Q(\jj\omega)}\neq0
\end{equation}
where nonsingularity follows from Lemma~\ref{lem:invertible}. Thus dynamic nodes 1 and 2 retain their baseline responses, while the nonzero oscillation at dynamic node 3 is eliminated. This establishes Pareto improvement at every positive frequency.
\end{IEEEproof}

\section{Case Studies}\label{sec:studies}

In this section, a 6-bus benchmark system and a modified IEEE 39-bus system were used to validate the infeasibility of single-actuator Pareto-improving control. The 6-bus system were further used to provide an example where multiple-actuator control achieving Pareto improvement. In both systems, SGs and VSMs were modeled as power sources. Simulations were conducted using PSS/E software\cite{siemens367}.

\subsection{6-bus Benchmark}

\subsubsection{Model}
\begin{figure}[!t]
\centering
\includegraphics{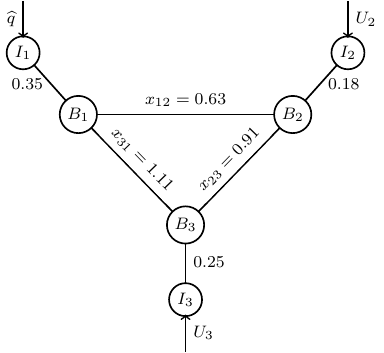}
\caption{6-bus benchmark system.}
\label{fig:simulation_network}
\end{figure}

Fig.~\ref{fig:simulation_network} shows the 6-bus system with two SGs and one VSM. Following Section~\ref{sec:model}, the retained dynamic buses are $\V=\{I_1,I_2,I_3\}$ and the eliminated algebraic buses are $\mathcal A=\{B_1,B_2,B_3\}$. The source is at dynamic bus $I_1$, and actuators are at dynamic buses $I_2$ and $I_3$. All six angle outputs are evaluated.

The system and device bases are 100 MVA, 230 kV, and 60 Hz. Reactances are labeled in per unit. The network is lossless and has no loads or shunts. The systems operates at unit voltages, zero angles, and zero power injections. Table~\ref{tab:simulation_parameters} lists the device parameters. The SGs use GENROU in its constant-internal-voltage, equal-reactance, zero-saturation limit. The VSM uses REGFM\_B1 model \cite{regfmb1} with direct damping $D_1=6.8$ and $T_{pf}=0$; additional droop and washout damping are disabled, while voltage control and current limits remain active.

\begin{table}[!t]
\caption{6-bus Benchmark Device Parameters}
\label{tab:simulation_parameters}
\centering
\begin{tabular}{lccc}
\hline
Device & $H_i$ (s) & $D_{g,i}$ & Source reactance (p.u.)\\
\hline
$I_1$: synchronous machine & 6.0 & 1.6 & 0.35\\
$I_2$: synchronous machine & 3.2 & 1.3 & 0.18\\
$I_3$: VSM & 6.4 & 6.8 & 0.25\\
\hline
\end{tabular}
\smallskip
\end{table}

At this operating point, edge weights equal to reciprocal reactances. Applying~\eqref{eq:network_reduction} gives $w_{12}=0.716480$, $w_{23}=0.696861$, and $w_{13}=0.449909$ in the reduced Laplacian $L$. The swing coefficients are $m_i=2H_i/(2\pi f_0)$ and $d_i=D_{g,i}/(2\pi f_0)$, with $f_0=60$ Hz. Thus the reduced network satisfies Theorem~\ref{thm:main}, and the six-output response matrix is
\begin{equation}\label{eq:simulation_transfer}
 F(p)=\begin{bmatrix}I\\T\end{bmatrix}
       (Mp^2+Dp+L)^{-1}
\end{equation}

\subsubsection{Single-Actuator Feasibility in the 6-bus Benchmark}

For actuator $I_c$, $c\in\{2,3\}$, let $v=U_c/\widehat q$ and
\begin{equation}\label{eq:simulation_normalized}
 \rho_i^{(c)}=\frac{F_{B_i,c}(\jj\omega)}{F_{B_i,1}(\jj\omega)},\qquad
 \frac{\widetilde Y_{B_i}}{\widetilde a_{B_i}}=1+\rho_i^{(c)}v
\end{equation}

Let $g$ be the largest gap between consecutive phases of $\rho_i^{(c)}$ around the circle, including the wraparound gap. The shortest arc containing these phases has width $360^\circ-g$, so the open-half-plane condition is equivalent to $g>180^\circ$. A gap $g\leq180^\circ$ rules out Pareto improvement across all six buses.

\begin{figure}[!t]
\centering
\includegraphics[width=\columnwidth]{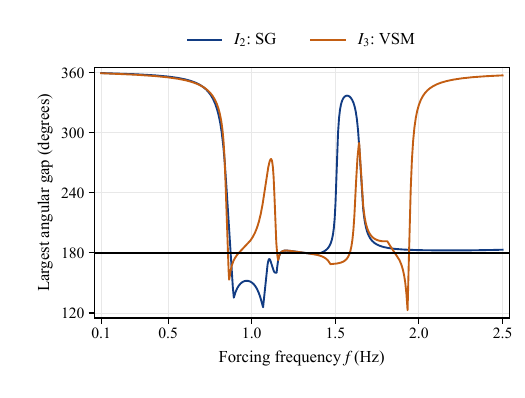}
\caption{Largest angular gap of normalized eliminated-bus responses in the 6-bus benchmark system.}
\label{fig:simulation_scan}
\end{figure}

Fig.~\ref{fig:simulation_scan} shows the largest gap results for the single-actuator cases with the SG in $I_2$ and the VSM in $I_3$, respectively. It could be observed that the simulation results agree with the theoretical analysis. When these two actuators operate individually, neither can achieve Pareto improvement over the entire frequency band, thereby confirming the validity of the analysis in this paper.

\subsubsection{Overcoming the Single-Actuator Limitation}
We select $\omega=7.27$ rad/s ($f\approx1.157$ Hz) for a further discussion, where the gaps are $171.801^\circ$ for $I_2$ and $172.921^\circ$ for $I_3$, both below $180^\circ$. Thus neither actuator alone can achieve Pareto improvement across all six outputs at this frequency. A disturbance of $q(t)=0.005\cos(7.27t)$ p.u., with a 0.5 MW amplitude after the ramp, is added to the oscillation source.

Table~\ref{tab:simulation_inputs} gives the control amplitudes normalized by $\widehat q=0.005$ p.u. For each single-actuator case, the control inputs are selected to achieve a 50\% reduction in oscillation amplitude at the corresponding eliminated bus. For multi-actuator control, the inputs at $I_2$ and $I_3$ are selected jointly to minimize the largest predicted amplitude ratio across all six buses.

\begin{table}[!t]
\caption{Control Phasors Normalized by the Source Amplitude}
\label{tab:simulation_inputs}
\centering
\begin{tabular}{lcc}
\hline
Configuration & $U_2/\widehat q$ & $U_3/\widehat q$\\
\hline
Uncontrolled & $0$ & $0$\\
Only $I_2$ & $0.851+\jj0.055$ & $0$\\
Only $I_3$ & $0$ & $0.418+\jj0.038$\\
Multi-actuator & $1.380+\jj0.082$ & $1.097-\jj0.103$\\
\hline
\end{tabular}
\end{table}

\begin{figure}[!t]
\centering
\includegraphics[width=\columnwidth]{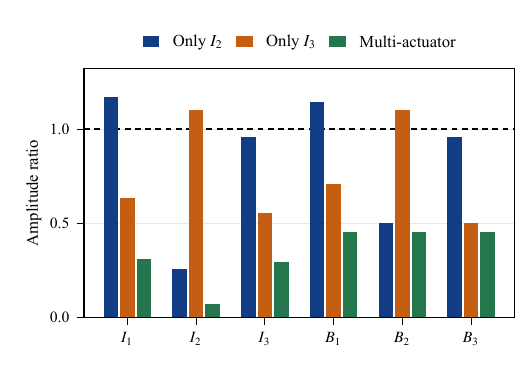}
\caption{Oscillation amplitude ratios at dynamic and eliminated buses under single-actuator and multi-actuator control.}
\label{fig:simulation_ratios}
\end{figure}

Fig.~\ref{fig:simulation_ratios} compares oscillation amplitudes at dynamic and eliminated buses under single-actuator and multi-actuator control. The bars show simulated amplitude ratios relative to the uncontrolled responses. And the dashed line at 1.0 marks the uncontrolled amplitude.

With $I_2$ alone, the $B_2$ amplitude ratio falls to $0.4998$, but $B_1$ rises to $1.1433$. With $I_3$ alone, $B_3$ falls to $0.5002$, but $B_2$ rises to $1.1012$. These single-actuator inputs reduce the targeted oscillation at the expense of another eliminated bus, illustrating the tradeoff. With multi-actuator control, amplitudes at all six buses decrease, and the largest measured ratio is $0.45571$. The coordinated inputs therefore achieve Pareto improvement, overcoming the single-actuator limitation at the selected frequency.

\begin{figure*}[!t]
\centering
\includegraphics[width=\textwidth]{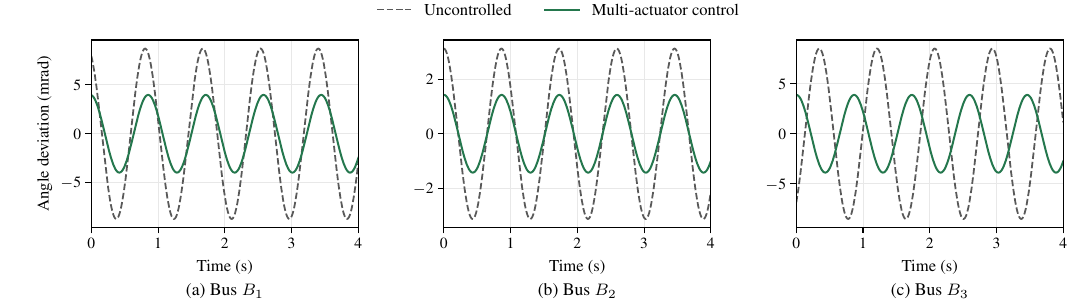}
\caption{Oscillation waveforms at eliminated buses with and without multi-actuator control.}
\label{fig:simulation_time}
\end{figure*}

Fig.~\ref{fig:simulation_time} shows the oscillations at buses $B_1$, $B_2$, and $B_3$ with and without multi-actuator control.

\subsection{Modified IEEE 39-Bus System}

\subsubsection{Model}
In this subsection, we further adopted a modified IEEE 39-Bus System~\cite{griddyn39} for testing, where SGs at buses 32, 34, and 36 were replaced with VSMs. Loads are modeled by constant impedances. The system uses a 100 MVA base and 50 Hz nominal frequency. VSM inertia constants are 4.0, 4.5, and 5.0 s, respectively. Each has $D_1=20$, coupling reactance 0.20 p.u., $T_{pf}=0.02$ s, and a steady-state current limit of 1.2 p.u. Additional droop and washout damping are disabled; voltage control and current limits remain active.

\begin{figure}[!t]
\centering
\includegraphics[width=\columnwidth]{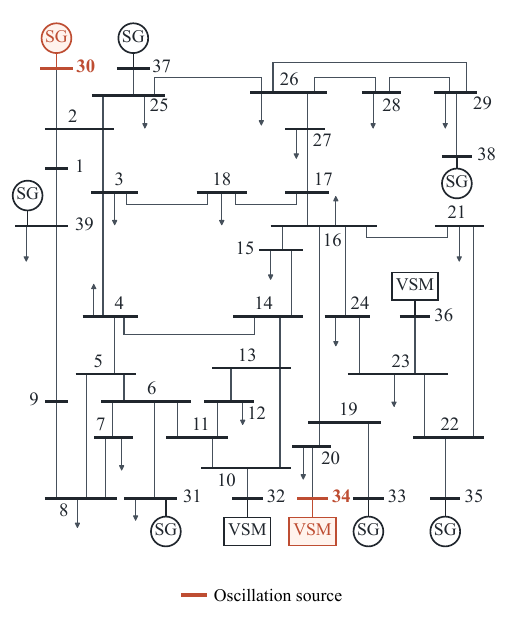}
\caption{Modified IEEE 39-bus system and oscillation source locations.}
\label{fig:ieee39_network}
\end{figure}

Fig.~\ref{fig:ieee39_network} shows the modified system, SG/VSM types, and the two oscillation source locations.

\subsubsection{Single-Actuator Feasibility in the modified IEEE 39-bus system}
The SG at bus 30 and the VSM at bus 34 are considered separately as oscillation sources. For each source, each of the other nine SGs and VSMs is considered in turn as the single damping actuator. All 39 bus angles are observed. For VSMs, the active-power references are modulated for source or actuator injections; for SGs, the TGOV1 references $R_{\mathrm{gov}}u$ are modulated, retaining governor dynamics. The scan covers 0.1--3 Hz at 0.05 Hz intervals. Each input location is simulated separately, and its fundamental response is normalized by the applied reference phasor.

\begin{figure*}[!t]
\centering
\includegraphics[width=\textwidth]{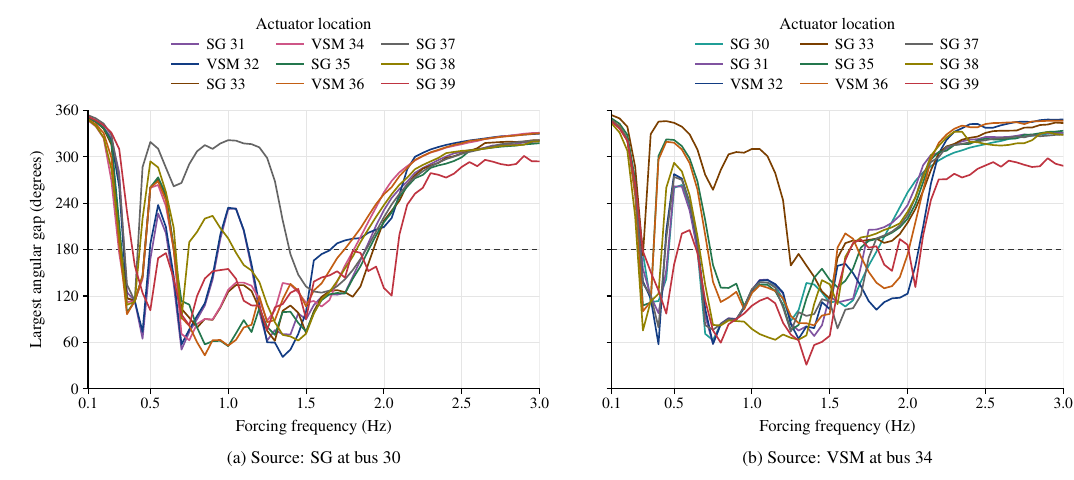}
\caption{Largest angular gap of normalized responses across all 39 bus angles for each source and its nine candidate actuators over 0.1--3 Hz.}
\label{fig:ieee39_scan}
\end{figure*}

Fig.~\ref{fig:ieee39_scan} shows nine actuator curves per source, with the dashed line marking $180^\circ$. All 18 gaps crosses this threshold in part. Hence none of the candidate single-actuator locations permits Pareto improvement for either source.

\section{Discussion}\label{sec:discussion}
\subsection{Implications for Power Systems}
Theorem~\ref{thm:main} establishes a limitation of single-actuator control in the model considered here, while the multi-actuator example illustrates how coordinated inputs can overcome it. These findings indicates that relying on a single actuator to suppress forced oscillations may have inherent limitations. They also suggest that multi-actuator architectures may be worth further investigation, particularly with respect to actuator placement and coordination. Such studies could help assess their potential for practical forced oscillation damping control.

The results also suggest care in interpreting reductions in oscillations measured at selected locations. With a persistent forcing source, a smaller measured frequency oscillation may only reflect local cancellation while responses at other nodes increase. Additional measurements or model-based response estimates could therefore complement the feedback signals and provide a broader view of the effects of control across the network.

\subsection{Limitations and Future Work}
The analysis is limited to small-signal active-power disturbances and control injections. Forced oscillations can also originate from excitation and reactive-power controls, as well as other converter mechanisms~\cite{nerc2017,chen2025}. Moreover, every retained dynamic node must admit the swing-equation representation used here. This covers synchronous machines and grid-forming controls with compatible dynamics, but does not generally describe grid-following converters or other grid-forming implementations.

Pareto improvement itself is a strict performance criterion: no nodal amplitude may increase, regardless of its initial severity. In practice, a modest increase at a bus with initially weak oscillations may be acceptable if it enables a substantial reduction in a severely affected area. Such a tradeoff can be valuable for system operation even though it does not qualify as a Pareto improvement.

Future work will focus on extending the feasibility analysis to reactive-power channels and more general converter dynamics, and on developing control criteria that account for oscillation severity and allow bounded local amplitude increases in exchange for greater reductions in severely affected areas.

\section{Conclusion}\label{sec:conclusion}
This work examined whether active-power damping control can reduce a forced oscillation without increasing its amplitude at any other nodes in the modeled network. Pareto improvement makes this requirement explicit for all dynamic and eliminated nodes. The necessary conditions and impossibility theorem show that, within the linear swing-network model, single-actuator control cannot meet this criterion at all positive forcing frequencies: every distinct source--actuator pair has at least one frequency at which no choice of control amplitude and phase is sufficient. Further, a three-node counterexample shows that multi-actuator control can overcome the single-actuator limitation. 

These results suggest that evaluating responses beyond the feedback locations can complement local measures of control performance. They also motivate further study of actuator placement and coordination for forced oscillation damping control.

\appendices
\section{Proof of the Cofactor Structure}\label{app:cofactors}
\subsection{Positive Definiteness and Diagonal Cofactors}
For the reduced-network Laplacian $L$, we show $L_{\mathcal J\mathcal J}\succ0$ whenever $\mathcal J\subset\V$ is nonempty and proper. Extend $x\in\C^{|\mathcal J|}$ by zeros outside $\mathcal J$ to form $\widetilde x$. Then
\begin{equation}\label{eq:groundenergy}
 x^*L_{\mathcal J\mathcal J}x=\sum_{\{i,k\}\in\E}w_{ik}|\widetilde x_i-\widetilde x_k|^2
\end{equation}
All summands are nonnegative. Since $w_{ik}>0$, a zero sum requires equal values at both ends of every edge. Paths connect every pair of dynamic nodes, so all entries are equal. An entry outside $\mathcal J$ is zero; hence $x=0$, proving positive definiteness.

Now set $\mathcal J=\V\setminus\{i\}$. Deleting row and column $i$ gives
\begin{equation}
 N_{ii}(p)=\det(M_{\mathcal J\mathcal J}p^2+D_{\mathcal J\mathcal J}p+L_{\mathcal J\mathcal J})
\end{equation}
The leading term is $\det(M_{\mathcal J\mathcal J})p^{2(n-1)}$, with positive coefficient. If $p_0$ is a root, the matrix is singular, so it sends some $x\neq0$ to zero. Left multiplication by $x^*$ gives
\begin{equation}
 (x^*M_{\mathcal J\mathcal J}x)p_0^2+(x^*D_{\mathcal J\mathcal J}x)p_0+x^*L_{\mathcal J\mathcal J}x=0
\end{equation}
These positive real coefficients give a negative root sum and a positive root product. Thus real roots are negative; conjugate roots have negative real part. Hence $\Rea p_0<0$, proving part 1 of Lemma~\ref{lem:cofactors}.

\subsection{Degree of the Source--Actuator Cofactor}
To find the degree of $N_{sc}$, write $Q=Z-W$, where $W_{ik}=w_{ik}$ for $i\neq k$, $W_{ii}=0$, and
\begin{equation}
 Z(p)=\diag(m_i p^2+d_i p+L_{ii})_{i=1}^n
\end{equation}
Since $Z^{-1}=O(|p|^{-2})$, large $|p|$ gives $\|Z^{-1}W\|<1$, allowing the inverse expansion
\begin{equation}\label{eq:neumann}
 H(p)=\sum_{k=0}^{\infty}(Z(p)^{-1}W)^kZ(p)^{-1}
\end{equation}
Each factor $W$ selects one edge. A contribution to entry $(s,c)$ along a walk $v_0=s,v_1,\ldots,v_k=c$ with $k$ edges is
\begin{equation}\label{eq:walkterm}
 \left(\prod_{h=0}^{k-1}w_{v_hv_{h+1}}\right)
 \left(\prod_{h=0}^{k}Z_{v_hv_h}(p)^{-1}\right)
\end{equation}
No walk exists for $k<\ell$. At $k=\ell$, each walk is a shortest path with $\ell+1$ distinct dynamic nodes. Using $Z_{ii}^{-1}=m_i^{-1}p^{-2}+O(|p|^{-3})$ gives
\begin{equation}\label{eq:leadingH}
 h_{sc}(p)=\kappa_{sc}p^{-2\ell-2}+O(|p|^{-2\ell-3})
\end{equation}
where
\begin{equation}\label{eq:kappa}
 \kappa_{sc}=\sum_{\mathcal P\in\mathfrak P_{sc}}
 \frac{\prod_{\{i,k\}\in\E(\mathcal P)}w_{ik}}
      {\prod_{i\in\V(\mathcal P)}m_i}>0
\end{equation}
Here $\mathfrak P_{sc}$ denotes the shortest paths, with sets of dynamic nodes and edges denoted by $\V(\mathcal P)$ and $\E(\mathcal P)$. Positive weights and inertias make $\kappa_{sc}>0$. The longer-walk terms together satisfy the norm bound
\begin{equation}
 \frac{\|Z^{-1}W\|^{\ell+1}\|Z^{-1}\|}{1-\|Z^{-1}W\|}
 =O(|p|^{-2\ell-4})
\end{equation}
Their faster decay cannot alter the leading term. Finally, $N_{sc}=h_{sc}\det Q$ and $\det Q(p)=\det(M)p^{2n}+O(|p|^{2n-1})$ yield
\begin{equation}
 \begin{split}
 N_{sc}(p)={}&\det(M)\kappa_{sc}p^{2(n-\ell-1)}\\
 &+O(|p|^{2(n-\ell-1)-1})
 \end{split}
\end{equation}
Thus $N_{sc}$ has degree $2(n-\ell-1)$ and positive leading coefficient, proving part 2.

\subsection{Common Constant Terms}
At $p=0$, $Q(0)=L$. Connectedness makes the solutions of $Lx=0$ exactly the constant vectors. The adjugate matrix $\adj L$, whose entries are $N_{ik}(0)$, satisfies
\begin{equation}
 L\adj L=(\adj L)L=0
\end{equation}
The two zero products show that every column and every row is constant. Thus all entries equal some $\tau$. A diagonal entry is $\det L_{\mathcal J\mathcal J}>0$ for $\mathcal J=\V\setminus\{i\}$, by positive definiteness. Hence $N_{ik}(0)=\tau>0$ for every pair, completing Lemma~\ref{lem:cofactors}.

\section{Proof of the Continuous Phase Formula}\label{app:phase}
Factor $\psi$ into real-root and conjugate-pair factors. No root lies on the imaginary axis. Normalize each factor to one at $p=0$; the remaining multiplier $\psi(0)>0$ does not affect phase.

For a real root $p_{\mathrm r}$, the factor at $p=\jj\omega$ is $1-\jj\omega/p_{\mathrm r}$. As $\omega$ grows, it moves upward from one for $p_{\mathrm r}<0$ and downward for $p_{\mathrm r}>0$. Its phase tends to $\pi/2$ or $-\pi/2$, respectively. Repeated roots repeat this contribution.

For a conjugate pair $\alpha\pm\jj\beta$, with $\beta>0$ and $\alpha\neq0$, the factor is
\begin{equation}
 \psi_{\mathrm{pair}}(p)=\frac{p^2-2\alpha p+\alpha^2+\beta^2}{\alpha^2+\beta^2}
\end{equation}
so
\begin{equation}
 \psi_{\mathrm{pair}}(\jj\omega)=1-\frac{\omega^2}{\alpha^2+\beta^2}
 -\jj\frac{2\alpha\omega}{\alpha^2+\beta^2}
\end{equation}
For $\alpha<0$, it stays above the real axis. The negative quadratic real term eventually dominates the linear imaginary term, so its phase approaches $\pi$. For $\alpha>0$, it stays below the axis and its phase approaches $-\pi$.

Each left-half-plane root therefore contributes $\pi/2$, and each right-half-plane root contributes $-\pi/2$. Count them as $z_-$ and $z_+$, including repetitions. Adding factor phases continuously from zero gives the phase of $\psi(\jj\omega)$. Since $z_-+z_+=d_\psi$, its limit is
\begin{equation}
 \frac{\pi}{2}z_- -\frac{\pi}{2}z_+
 =\frac{\pi}{2}(d_\psi-2z_+)
\end{equation}
Any other continuous phase differs by an integer multiple of $2\pi$. Continuity makes that integer constant. The same initial value forces the difference to be zero, proving uniqueness and Lemma~\ref{lem:phase}.

\section{Control Construction for All Nodes}\label{app:full_node_control}
Fix $\omega>0$ and use the full-node responses in~\eqref{eq:full_output_response}. The affected set $\widetilde{\I}=\{i\in\mathcal N:\widetilde b_i\neq0\}$ is nonempty because $\widetilde b_c=h_{cc}(\jj\omega)\neq0$ by Lemma~\ref{lem:invertible}.

If any affected node has $\widetilde a_i=0$, every nonzero input creates an oscillation there, ruling out Pareto improvement. Otherwise, define $\widetilde r_i=\widetilde b_i/\widetilde a_i$ on $\widetilde{\I}$. Improvement is possible exactly when one input direction has an opposing component at every affected node:
\begin{equation}\label{eq:full_node_direction}
 \Rea(\widetilde r_iU_{\mathrm{dir}})<0
 \qquad\text{for every }i\in\widetilde{\I}
\end{equation}
Equivalently, these ratios lie in one open half-plane whose boundary passes through the origin.

For necessity, any improving input is nonzero. Expanding $|1+\widetilde r_iU|\leq1$ gives $\Rea(\widetilde r_iU)\leq-|\widetilde r_iU|^2/2<0$ at every affected node. Thus $U_{\mathrm{dir}}=U$ satisfies~\eqref{eq:full_node_direction}.

For sufficiency, scale a suitable direction by $U=\varepsilon_U U_{\mathrm{dir}}$, keeping its phase fixed. Choose
\begin{equation}\label{eq:stepsize}
 \begin{aligned}
 0&<\varepsilon_U<\varepsilon_{U,\max}(U_{\mathrm{dir}}),\\
 \varepsilon_{U,\max}(U_{\mathrm{dir}})
 &=\min_{i\in\widetilde{\I}}
   \frac{-2\Rea(\widetilde r_iU_{\mathrm{dir}})}{|\widetilde r_iU_{\mathrm{dir}}|^2}
 \end{aligned}
\end{equation}
Each node's bound is finite and positive by~\eqref{eq:full_node_direction}. Their minimum is positive because the affected set is finite and nonempty. Below this bound, the negative linear term dominates the positive quadratic term:
\begin{equation}
 \begin{aligned}
 |1+\varepsilon_U\widetilde r_iU_{\mathrm{dir}}|^2-1
 &=2\varepsilon_U\Rea(\widetilde r_iU_{\mathrm{dir}})\\
 &\quad+\varepsilon_U^2|\widetilde r_iU_{\mathrm{dir}}|^2<0
 \end{aligned}
\end{equation}
Hence every affected amplitude decreases. At other nodes, $\widetilde b_i=0$ leaves the response unchanged. At least one node is affected, so this input satisfies the full-node Pareto criterion~\eqref{eq:target}.

\bibliographystyle{IEEEtran}
\bibliography{ref}
\end{document}